\documentclass[british]{tMOP2e}
\usepackage[T1]{fontenc}
\usepackage[latin1]{inputenc}
\usepackage{color}
\usepackage{amsmath}
\usepackage{graphicx}
\usepackage{amssymb}

\makeatletter
\@ifundefined{textcolor}{}
{%
 \definecolor{BLACK}{gray}{0}
 \definecolor{WHITE}{gray}{1}
 \definecolor{RED}{rgb}{1,0,0}
 \definecolor{GREEN}{rgb}{0,1,0}
 \definecolor{BLUE}{rgb}{0,0,1}
 \definecolor{CYAN}{cmyk}{1,0,0,0}
 \definecolor{MAGENTA}{cmyk}{0,1,0,0}
 \definecolor{YELLOW}{cmyk}{0,0,1,0}
 }

\usepackage{color}
\usepackage{soul}
\usepackage{dcolumn}
\usepackage{bm}
\usepackage{rotating}
\usepackage{psfrag}
\usepackage{babel}

\makeatother

\usepackage{babel}

\begin{document}

\title{Classical diffusive dynamics for the quasiperiodic kicked rotor}

\author{Gabriel Lemarié$^{a}$, Dominique Delande$^{b}$, Jean Claude Garreau$^{c}$ and Pascal Szriftgiser$^{c}$$^{\ast}$\thanks{$^\ast$Corresponding author. Email: Pascal.Szriftgiser@univ-lille1.fr
\vspace{6pt}}\\\vspace{6pt} 
$^{a}${\em{Service de Physique de l'État Condensé (CNRS URA 2464),\\ IRAMIS/SPEC, CEA Saclay, F-91191 Gif-sur-Yvette, France}};\\
$^{b}${\em{Laboratoire Kastler Brossel, UPMC-Paris 6, ENS, CNRS; 4 Place Jussieu,\\
F-75005 Paris, France}};\\
$^{c}${\em{Laboratoire de Physique des Lasers, Atomes et Mol{é}cules, Universit{é}
Lille 1 Sciences et Technologies, UMR CNRS 8523; CERLA; F-59655 Villeneuve
d'Ascq Cedex, France}};}

\maketitle

\begin{abstract}
We study the \emph{classical} dynamics of a quasiperiodic kicked rotor, whose \emph{quantum}
counterpart is known to be an equivalent of the 3D Anderson model. Using
this correspondence allowed for a recent experimental observation of
the Anderson transition with atomic matter waves. In such a context, it is
particularly important to assert the chaotic character of the classical
dynamics of this system. We show here that it is a 3D anisotropic diffusion. Our simple analytical predictions for
the associated diffusion tensor are found in good agreement with the results
of numerical simulations.
\end{abstract}


\section{Introduction}

One of the most remarkable interference effects is
certainly the Anderson localisation \cite{Anderson:LocAnderson:PR58}:
A classical random walker displays diffusive dynamics. But a quantum-coherent
one (i.e. which keeps a memory of its phase) can either diffuse,
or localise. This depends on the relative importance
of diffusive transport and interference effects.

Most studied wave systems which may show localisation are those where a wave propagates in a disordered medium, where
disorder induces a random walk. However, the phenomenon of random
walk is not restricted to disordered systems, as it can occur in very simple chaotic systems
such as the well-known kicked rotor \cite{Chirikov:ChaosClassKR:PhysRep79}, described
by the Hamiltonian: 
\begin{equation}
H=\dfrac{p^{2}}{2}+K\cos x\sum_{n}\delta(t-n)\;,\label{eq:HKR}
\end{equation}
where $p$ is the momentum conjugated to the variable $x$ (which,
as it appears only in the argument of a cosinus, can be restricted
to $[0,2\pi]$), the particle mass is taken equal to unity and $K$
represents the amplitude of the sinusoidal potential.
This system has one degree of freedom, but the
presence of a temporal forcing and its non-linear character
allow the existence of a chaotic regime. More specifically, for
values of the so-called stochasticity parameter $K\gtrsim4$, the classical dynamics of that
system along $p$ is a random walk: the walker {}`` particle $p$''
diffuses \cite{Chirikov:ChaosClassKR:PhysRep79}.

In 1995, Raizen and coworkers realised a kicked rotor by exposing laser-cooled
atoms to a pulsed standing wave \cite{Raizen:QKRFirst:PRL95}, which
is {}``seen'' by the atoms as a sinusoidal potential acting on the
centre of mass variables. However, for cold enough atoms, the
de Broglie wavelength becomes comparable to the spatial period of the standing wave, in which case their dynamics is quantum
(wavelike) and one can observe the phenomenon of ``dynamical localisation'' \cite{Casati:LocDynFirst:LNP79,Raizen:QKRFirst:PRL95},
that is, the freezing
of the diffusive transport of the walker in the
momentum space. This phenomenon has been shown to
be equivalent to the Anderson localisation in 1D \cite{Fishman:LocDynAnderson:PRA84, Atland:SuperSymKR:PRL96, Atland:DiagKR:PRL93}.
An impressive surge of theoretical and experimental work on the subject
followed \cite{Raizen:LDynNoise:PRL98,DArcy:AccModes:PRL99,Philips:QuantumResTalbot:PRL99,Monteiro:DoubleKick:PRL04,Renzoni:RatchetOptLat:PRL04,Summy:QKAPhaseSpace:PRL06,Gong:HarperKR:PRA08,GarciaGarcia:AndersonKR3D:PRE09,AP:Bicolor:PRL00,AP:SubFourier:PRL02,AP:Reversibility:PRL05,AP:SubFMecs:EPL05,AP:PetitPic:PRL06}.

The correspondence between the kicked-rotor and the
Anderson model can be extended to higher dimensions: It turns out
that a quasiperiodic kicked rotor with $d$ incommensurable temporal
frequencies is equivalent to a $d$ dimensional Anderson model \cite{Casati:IncommFreqsQKR:PRL89, Basko:WeakDynLocQP:PRL03, Lemarie:UnivAnderson:EPL09}.
Adding two new (incommensurate) frequencies to the Hamiltonian (\ref{eq:HKR})
thus prompts to the observation of the \emph{Anderson
transition}, which exists only in 3 (or higher)
dimensions. We have engineered in our cold atom experiment \cite{AP:Anderson:PRL08,AP:AndersonLong:PRA09}
a quasiperiodic realisation of the kicked rotor described by the Hamiltonian
\begin{equation}
H_{\text{qp}}=\dfrac{p^{2}}{2}+K\cos x[1+\varepsilon\cos(\omega_{2}t)\cos(\omega_{3}t)]\sum_{n}\delta(t-n)\;.\label{eq:Hqp}
\end{equation}
to study and characterise this transition, including, in particular,
the first non-ambiguous measurement of its critical exponent. Here,
$\varepsilon$ is the amplitude of the quasi-periodic modulation of
the strength of the kicks, and the frequencies $\omega_{2}$ and $\omega_{3}$
must satisfy a condition of incommensurability. 

It can be shown \cite{Shepelyanky:Kq:PD87,Casati:IncommFreqsQKR:PRL89,AP:AndersonLong:PRA09} that the above 1D system (\ref{eq:Hqp}) has the same transport properties as that of the following 3D kicked ``rotor'':
\begin{align}
H_{3}=  \dfrac{p_{1}^{2}}{2}+\omega_{2}p_{2}+\omega_{3}p_{3}+K\cos x_{1}[1+\varepsilon\cos x_{2}\cos x_{3}]\sum_{n}\delta(t-n)\;,\label{eq:KR3Dquasiper}
\end{align}
where $\boldsymbol{p}=(p_{1},p_{2},p_{3})$ is the momentum conjugated
to $\boldsymbol{x}=(x_{1},x_{2},x_{3})$. In fact, the temporal evolution of an arbitrary initial
condition $\psi_{\text{qp}}(\theta,t=0)$ for $H_{\text{qp}}$ (\ref{eq:Hqp}) is identical to that of the corresponding $\psi_{3}$ : \begin{equation}
\psi_{3}(\boldsymbol{\theta},t=0)=\psi_{\text{qp}}(\theta_{1},t=0)\delta(\theta_{2})\delta(\theta_{3})\;,\end{equation}
for $H_3$ (\ref{eq:KR3Dquasiper}) (see \cite{AP:AndersonLong:PRA09}). Thus, our
experience with the quasiperiodic kicked rotor can be regarded as
an experience of transport of waves in a 3D medium originating from
a time-pulse $\delta(t=0)$ of a plane source (invariant along the
directions $p_{2}$ and $p_{3}$), where transport properties are
actually observed in the direction perpendicular to the emitter. This
presents strong similarities with the methods of study of transport
of classical waves in 3D disordered media \cite{Maret:AndersonTransLight:PRL06}.

Note
that (\ref{eq:KR3Dquasiper}) is not the
standard 3D kicked rotor \cite{GarciaGarcia:AndersonKR3D:PRE09}:
\begin{align}
\tilde{H}_{3}=  \dfrac{p_{1}^{2}}{2}+\omega_{2}\dfrac{p_{2}^{2}}{2}+\omega_{3}\dfrac{p_{3}^{2}}{2}+
K\cos x_{1}\cos x_{2}\cos x_{3}\sum_{n}\delta(t-n)\;.\label{eq:KR3D}
\end{align}
Two characteristics of (\ref{eq:KR3Dquasiper}) are noteworthy:
First, the linear dependence of the transverse kinetic energy in $p_{2}$
and $p_{3}$. Second, the {}``true'' spatial direction
{}``1'' and the transverse ``virtual'' directions
{}``2'' and {}``3'' present an anisotropy controlled
by the parameter $\varepsilon$. How these features affect the classical dynamics of (\ref{eq:KR3Dquasiper}) as compared to that
isotropically diffusive of (\ref{eq:KR3D}) is the subject of this study. 

Our motivations are the
following. The linear dispersion of the transverse
degrees of freedom in the Hamiltonian (\ref{eq:KR3Dquasiper}) seems,
at first glance, to imply
that only the direction {}``1'' in (\ref{eq:KR3Dquasiper}) is
disordered while the transverse directions {}``2'' and {}``3''
are quasiperiodic (see \cite{Fishman:LocDynAnders:PRL82,Shepelyanky:Kq:PD87,GarciaGarcia:AndersonKR3D:PRE09,AP:AndersonLong:PRA09}).
This quasiperiodic character
could prevent the classical transport in the transverse directions, thus making the localisation observed
in the quantum regime not a consequence of subtle interference effects
but attributable to the peculiar
classical dynamics of this system \cite{Leboeuf:LocBicPot:PRA10}. This is however
not the case. The classical dynamics of (\ref{eq:KR3Dquasiper}) is, as we shall see, fully diffusive in all
directions, and the localisation observed in the quantum regime is therefore
a quantum effect of the same nature as Anderson
localisation. Second, the anisotropy of the
classical transport is known to greatly affect the properties of the
localisation, including a non-negligible dependence of the transition
threshold on the degree of anisotropy (e.g. on the parameter $\varepsilon$) \cite{Evangelou:LocDelocSolids:PRB94}. We indeed
observed such a dependence (see the phase diagram in \cite{AP:AndersonLong:PRA09}).

\section{Analysis of the diffusive dynamics\label{sec:AnalysisDynamics}}

The Hamiltonian (\ref{eq:KR3Dquasiper}) being
periodic in time, its classical dynamics can be
studied by adopting a stroboscopic point of view, e.g. by focusing
on the system state just after the kicks. Integrating Hamilton's
equations from a kick to the following one, we obtain the equivalent
of the Standard Map \cite{Chirikov:ChaosClassKR:PhysRep79} for this system, which relates values of the
space and momentum vectors ($\boldsymbol{x}_{n}$ $\boldsymbol{p}_{n}$)
just after the $n_{th}$ kick to those $(\boldsymbol{x}_{n+1},\boldsymbol{p}_{n+1})$
after a period: \begin{align}
p_{1_{n+1}} & =p_{1_{n}}+K\sin\theta_{1_{n}}(1+\varepsilon\cos x_{2_{n}}\cos x_{3_{n}})\;,\nonumber \\
p_{2_{n+1}} & =p_{2_{n}}+K\varepsilon\cos x_{1_{n}}\sin x_{2_{n}}\cos x_{3_{n}}\;,\nonumber \\
p_{3_{n+1}} & =p_{3_{n}}+K\varepsilon\cos x_{1_{n}}\cos x_{2_{n}}\sin x_{3_{n}}\;,\label{eq:KR3DApplicationStandard}\\
x_{1_{n+1}} & =x_{1_{n}}+p_{1_{n+1}}\;,\nonumber \\
x_{2_{n+1}} & =x_{2_{n}}+\omega_{2}\;,\nonumber \\
x_{3_{n+1}} & =x_{1_{n}}+\omega_{3}\;.\nonumber
\end{align}
Starting from an initial condition ($\boldsymbol{x}_{0},\boldsymbol{p}_{0}$),
one can easily simulate the evolution of this system. In particular,
we want to see how a state initially localised at $\boldsymbol{p}_{0}=0$
expands as time passes. Afterwards, only properties averaged over
many initial conditions $\boldsymbol{p}_{0}=0$ and $\boldsymbol{x}_{0}\in[0,2\pi)$
will be considered. 

Note that
the temporal dependency (\ref{eq:KR3DApplicationStandard}) of the transverse variables $x_{2}$ and $x_{3}$
is trivial, and very
different from that obtained from (\ref{eq:KR3D}): 
\begin{align}
x_{2_{n+1}} & =x_{2_{n}}+\omega_{2}p_{2}\;,\nonumber \\
x_{3_{n+1}} & =x_{1_{n}}+\omega_{3}p_{3}\;.\label{eq:x2x3KR3D}
\end{align}
However, from (\ref{eq:KR3DApplicationStandard})
we can express ${p_{2}}_{n+1}$ as: \begin{equation}
{p_{2}}_{n+1}={p_{2}}_{0}+K\varepsilon\sum_{j=1}^{n}\cos x_{1_{j}}\sin(x_{2_{0}}+\omega_{2}j)\cos(x_{3_{0}}+\omega_{3}j)\;,\end{equation}
which shows that the chaotic character of the motion
in direction {}``1'' progressively {}``diffuses'' into the transverse
directions, which hints for a complex dynamics in all directions. Indeed, as in the case of the 1D kicked rotor,
the variables $x_{1_{j}}$ can be considered as uncorrelated random
variables, uniformly distributed over $[0,2\pi)$ (the quasi-linear
approximation is valid for large $K$) and $p_{2}$ thus
also performs a random walk \cite{Shepelyanski:Bicolor:PD83}. 

The randomisation of $x_{1}$ is due
to two conjugated phenomena: the sensitivity on initial conditions
and the folding of the phase-space. The sensitivity
on initial conditions is characterised by the Kolmogorov-Sinaï entropy
$h$ (an average of individual Lyapounov exponents over a chaotic part or all of the phase space - see \cite{Chirikov:ChaosClassKR:PhysRep79}), which gives the time-scale $1/h$ over which two trajectories initially close have diverged
by $\delta x_{1}\sim2\pi$.
The quasi-periodic action of $x_{2}$ and $x_{3}$ is to increase
the entropy $h$, as compared to the standard case of the kicked rotor.
Thus, we can say that the variable $x_{1}$ is at least as random
and uncorrelated as the variable $x$ of the kicked rotor \cite{Shepelyanski:Bicolor:PD83}. Indeed,
we know that, for $K\gtrsim4$, the whole phase-space of the kicked
rotor is chaotic and that $x$ can be considered as an uncorrelated
(or shortly correlated, the correlation time being $\sim1/h$) variable.
This will be true also for (\ref{eq:KR3DApplicationStandard}). However, for $\varepsilon$
not too small, the threshold in $K$ should be smaller. We have not
dealt with this issue of the threshold of the diffusive transport
in the 3D kicked ``rotor'' (\ref{eq:KR3Dquasiper}). In the following, we will restrict ourselves
to values of $K$ larger than $4$.

The dynamics of the system in momentum space is thus diffusive, characterised
by a diffusion tensor $D_{ij}\ (i,j=1,2,3)$ defined as:
\begin{equation}
D_{ij} = \text{lim}_{n\to\infty}\ \frac{\langle{p_{i}}_{n}{p_{j}}_{n}\rangle}{n}
\end{equation}
where $\langle X\rangle$ represents the average of $X$ over initial
conditions. We can evaluate approximately this diffusion tensor. Let
us consider the behaviour of $\langle{p_{2}}^{2}\rangle$, for example.
We can write this quantity as: \begin{align}
\langle{{p_{2}}_{n}}^{2}\rangle=  K^{2}\varepsilon^{2}\sum_{j,m=1}^{n-1}&\langle\cos{x_{1}}_{j}\cos{x_{1}}_{m}\rangle_{{x_{1}}_{0}} \times\langle\sin(x_{2_{0}}+\omega_{2}j)\sin(x_{2_{0}}+\omega_{2}m)\rangle_{{x_{2}}_{0}}\times\nonumber \\
 & \times\langle\cos(x_{3_{0}}+\omega_{3}j)\cos(x_{3_{0}}+\omega_{3}m)\rangle_{{x_{3}}_{0}}\;.\label{eq:moyp2}\end{align}
The correlation function $\langle\cos{x_{1}}_{j}\cos{x_{1}}_{m}\rangle_{{x_{1}}_{0}}$
decreases exponentially in $\vert j-m\vert$, as in the case of the
1D kicked rotor \cite{Shepelyanski:Bicolor:PD83}. So the main contribution
to the sum in (\ref{eq:moyp2}) is given by $j=m$. Performing the
average over the initial conditions, we finally obtain: \begin{align}
\langle{{p_{2}}_{n}}^{2}\rangle\approx\dfrac{K^{2}\varepsilon^{2}}{8}\times{n}\;,\end{align}
 in the quasi-linear approximation. Following those lines in the case
of the other quantities $\langle{p_{i}}{p_{j}}\rangle$, we see easily
that the diffusion tensor is diagonal and reads: \begin{align}
D_{11} & \approx(K^{2}/2)(1+\varepsilon^{2}/4)\;,\label{diff:tensorD11}\\
D_{22} & \approx K^{2}\varepsilon^{2}/8\;,\label{diff:tensorD22}\\
D_{33} & \approx K^{2}\varepsilon^{2}/8\;,\label{diff:tensorD33}\\
D_{i\neq j} & \approx0\;.\label{diff:tensorDij}\end{align}
 From this, it is clear that the diffusive transport is anisotropic
and that the axes {}``1'', {}``2'' and {}``3'' are the principal
axes.

\section{Confrontation with the results of numerical simulations}

\begin{figure}
\begin{centering}
\psfrag{a}{} \psfrag{x}{$n$} \psfrag{y}{ $\langle{p_{i}}_{n}{p_{j}}_{n}\rangle$}
\includegraphics[width=6cm]{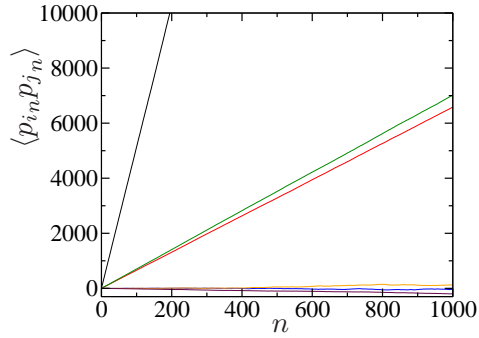} 
\par\end{centering}

\caption{\label{fig:diff1} Time evolution of the correlation
of classical momenta $\langle{p_{i}}{p_{j}}\rangle$. The non-diagonal
terms $i\neq j$ (orange, blue and maroon) keep small values all along the evolution,
whereas the diagonal terms increase linearly with time, the slope
of $\langle{p_{1}}^{2}\rangle$ (represented in black) being much
larger than the slopes of $\langle{p_{2}}^{2}\rangle$ (in red) and
$\langle{p_{3}}^{2}\rangle$ (in green). This shows that the dynamics
is anisotropically diffusive and that {}``1'', {}``2''
and {}``3'' are its principal axes. The dynamics
is approximately the same along the axes {}``2'' and {}``3'' (see
figure \ref{fig:D22D33vsKeps08}). The parameters are: $K=10$, $\varepsilon=0.8$,
$\omega_{2}=2\pi\sqrt{5}$ and $\omega_{3}=2\pi\sqrt{13}$.}

\end{figure}

\begin{figure}
\begin{centering}
\includegraphics[width=6cm]{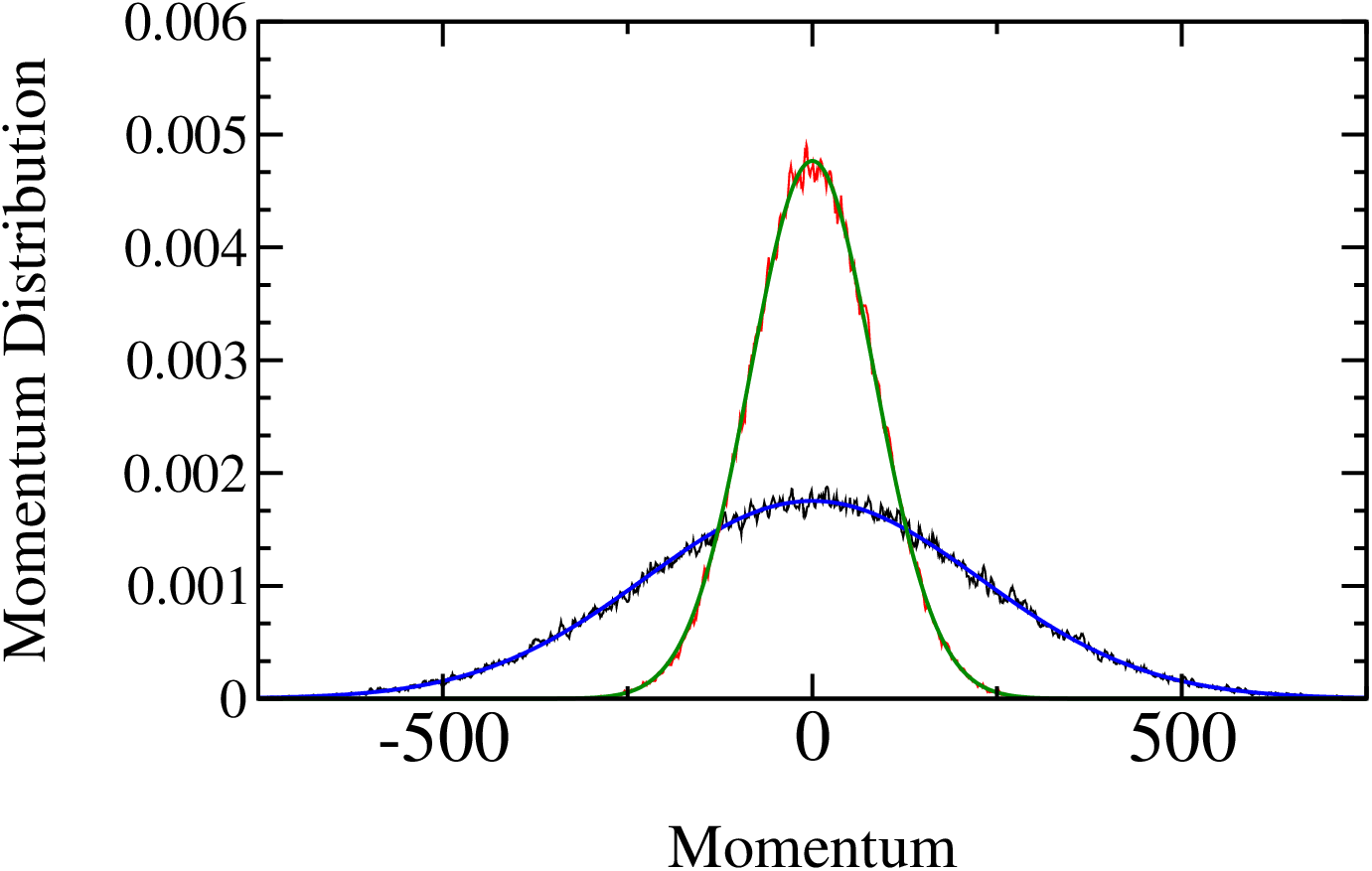} 
\par\end{centering}

\caption{\label{fig:diff2} Final momentum distributions along $p_{1}$ (in
black) and $p_{2}$ (in red), the distribution along $p_{3}$ being
approximately identical to that along $p_{2}$. After $1000$ kicks,
they display all a Gaussian shape characteristic of a diffusive motion.
The blue and green curves are fits by a Gaussian which do not show
any statistically significant deviation. Parameters are $K=10$, $\varepsilon=0.8$,
$\omega_{2}=2\pi\sqrt{5}$ and $\omega_{3}=2\pi\sqrt{13}$.}

\end{figure}

In this section we will compare the
predictions of the previous sections with numerical simulations.
First, we verify that the transport is indeed diffusive in all directions
(for $K\gtrsim4$), and that the non-diagonal coefficients $D_{i\neq j}$
of the diffusion tensor vanish. Figure \ref{fig:diff1} represents
the temporal evolution of $\langle{p_{i}}{p_{j}}\rangle$ for the
Hamiltonian (\ref{eq:KR3Dquasiper}) with $\omega_{2}=2\pi\sqrt{5}$
and $\omega_{3}=2\pi\sqrt{13}$ incommensurate with each other and
with $2\pi$ (this condition of incommensurability corresponds to
the case of the quasiperiodic kicked rotor), $K=10$ and $\varepsilon=0.8$.
The time evolution of the diagonal terms $\langle{p_{i}}^{2}\rangle$
is linear while the non-diagonal terms $\langle{p_{i}}{p_{j\neq i}}\rangle$
stay
very close to zero. Also, the distribution of each $p_{i}$ has a
Gaussian shape, characteristic of a diffusive transport (see figure
\ref{fig:diff2}).

\begin{figure}
\begin{centering}
\psfrag{a}{} \psfrag{x}{$K$} \psfrag{y}{ $D_{11}$} \includegraphics[width=6cm]{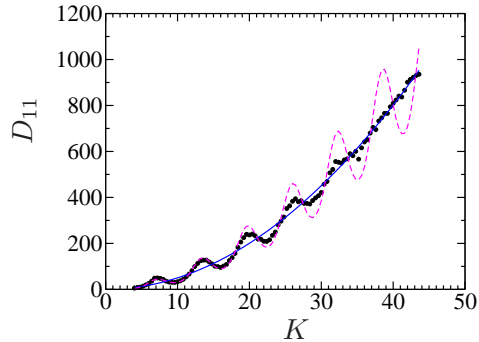} 
\par\end{centering}

\caption{\label{fig:D11vsKeps01} Dependence on $K$ of the longitudinal diffusion
coefficient $D_{11}$ (black points) when the anisotropy is strong:
$\varepsilon=0.1$. One observes oscillations around
the average behaviour equation (\ref{diff:tensorD11}) (represented
in blue) which follow the known corrections of the kicked rotor, equation (\ref{eq:oscilDKR})
(dashed magenta curve), but only for values of $K$ not too large.
When $K$ is increased, these oscillations decrease faster than in
the case of the kicked rotor. The parameters are the following: $\omega_{2}=2\pi\sqrt{5}$
and $\omega_{3}=2\pi\sqrt{13}$.}

\end{figure}

The dependence of the diffusion tensor versus $K$ is now studied
for $\varepsilon=0.1$, when the anisotropy is strong, and for $\varepsilon=0.8$,
when the anisotropy is much weaker.

For $\varepsilon=0.1$, $D_{11}(K)$ is found to oscillate around
its average behaviour, equation (\ref{diff:tensorD11}), the amplitude
of these oscillations decreasing as $K$ increases. This recalls the
oscillations of the diffusion coefficient of the 1D kicked rotor when
$K$ is varied. In the latter case, these oscillations appear due
to the presence of residual correlations $\langle\sin x_{n}\sin x_{0}\rangle$.
Taking into account correlations up to four periods, it leads to the
following approximation for the oscillations of $D(K)$ for the kicked
rotor \cite{Rechester:KRDiffCoeff:PRA81}:
\begin{equation}
D(K)\approx\dfrac{K^{2}}{2}\{1-2J_{2}(K)[1-J_{2}(K)]\}\;,\label{eq:oscilDKR}\end{equation}
 where $J_{2}(K)$ is the Bessel function of second order. In the
case of the Hamiltonian (\ref{eq:KR3Dquasiper}), we expect a similar
phenomenon to occur when $\varepsilon$ is small. $D_{11}(K)$ should
follow the same oscillations of $D(K)$ (\ref{eq:oscilDKR}). This
is indeed the case, but only for not too large values of $K$, as
can be seen in figure \ref{fig:D11vsKeps01}. As $K$ increases further,
the oscillations of $D_{11}(K)$ are quickly killed and we recover
the average behaviour equation (\ref{diff:tensorD11}).

\begin{figure}
\begin{centering}
\psfrag{b}{} \psfrag{x}{$K$} \psfrag{y}{ $D_{22}$, $D_{33}$}
\includegraphics[width=6cm]{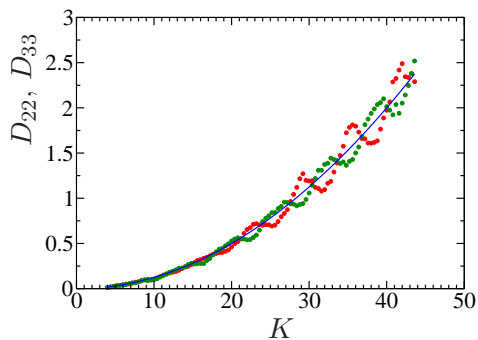} 
\par\end{centering}

\caption{\label{fig:D22D33vsKeps01} Dependence on $K$ of the transverse diffusion
coefficients $D_{22}$ (red points) and $D_{33}$ (green points) when
the anisotropy is strong: $\varepsilon=0.1$. The diffusion coefficients
are found to oscillate around their average behaviour, equations (\ref{diff:tensorD22})
and (\ref{diff:tensorD33}). The oscillating corrections do not decrease
in amplitude as $K$ increases, and they have the same period of the
oscillations of $D_{11}(K)$. The parameters are: $\omega_{2}=2\pi\sqrt{5}$
and $\omega_{3}=2\pi\sqrt{13}$.}

\end{figure}

\begin{figure}
\begin{centering}
\psfrag{b}{} \psfrag{x}{$K$} \psfrag{y}{ $D_{22}$} \includegraphics[width=6cm]{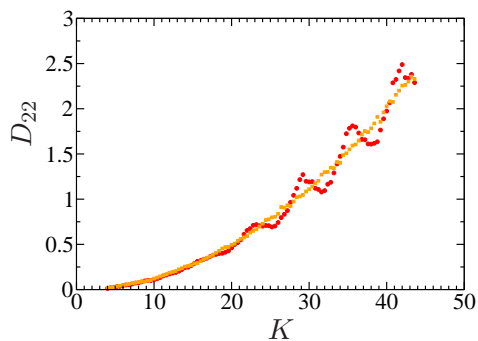} 
\par\end{centering}

\caption{\label{fig:D22KR3DvsKeps01} Comparison between the behaviour of $D_{22}$
vs $K$ for the Hamiltonian (\ref{eq:KR3Dquasiper}) (red points - see figure \ref{fig:D22D33vsKeps01})
and for the Hamiltonian (\ref{eq:KR3D}) where $x_{2}$ and $x_{3}$
follow equations (\ref{eq:x2x3KR3D}) (orange points). In the latter case, the oscillating
corrections, which are a signature of residual correlations between
kicks, decrease in amplitude as $K$ increases. The parameters are:
$\varepsilon=0.1$, $\omega_{2}=2\pi\sqrt{5}$ and $\omega_{3}=2\pi\sqrt{13}$.}

\end{figure}

The evolutions of the transverse diffusion coefficients $D_{22}$
and $D_{33}$ vs $K$ for $\varepsilon=0.1$ are represented in figure
\ref{fig:D22D33vsKeps01}. The oscillating corrections to the average
behaviour equations (\ref{diff:tensorD22}) and (\ref{diff:tensorD33})
do not seem to decrease in amplitude. On the contrary, in the case
of a true 3D anisotropic kicked rotor, these oscillations are found
to decrease in amplitude (see figure \ref{fig:D22KR3DvsKeps01}).
The trivial dependence of $x_{2}$ and $x_{3}$ in the case of the
Hamiltonian (\ref{eq:KR3Dquasiper}) is certainly responsible for this.

\begin{figure}
\begin{centering}
\psfrag{a}{} \psfrag{x}{$K$} \psfrag{y}{ $D_{11}$} \includegraphics[width=6cm]{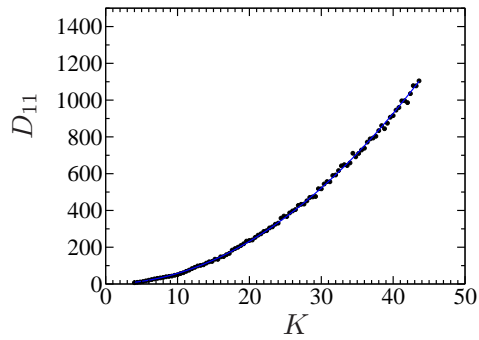} 
\par\end{centering}

\caption{\label{fig:D11vsKeps08} Dependence on $K$ of $D_{11}$ (black points)
when the anisotropy is small: $\varepsilon=0.8$. $D_{11}$ closely
follows the average behaviour prediction, equation (\ref{diff:tensorD11}),
represented in blue. In this case, correlations between kicks are
all negligible. The parameters are: $\omega_{2}=2\pi\sqrt{5}$ and
$\omega_{3}=2\pi\sqrt{13}$.}

\end{figure}

\begin{figure}
\begin{centering}
\psfrag{b}{} \psfrag{x}{$K$} \psfrag{y}{ $D_{22}$, $D_{33}$}
\includegraphics[width=6cm]{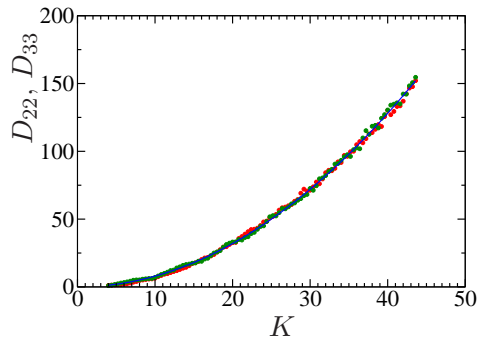} 
\par\end{centering}

\caption{\label{fig:D22D33vsKeps08} Dependence on $K$ of $D_{22}$ (red points)
and $D_{33}$ (green points) when the anisotropy is small: $\varepsilon=0.8$.
The diffusion coefficients closely follow the average behaviour predictions,
equations (\ref{diff:tensorD22}) and (\ref{diff:tensorD33}), represented
in blue. The parameters are: $\omega_{2}=2\pi\sqrt{5}$ and $\omega_{3}=2\pi\sqrt{13}$.}

\end{figure}

When the anisotropy is weak, $\varepsilon=0.8$, the residual correlations
between kicks, responsible for the oscillating corrections on the
diffusion tensor discussed above, are killed even for small $K$.
This is clearly seen in figures \ref{fig:D11vsKeps08} and \ref{fig:D22D33vsKeps08}
where the diffusion tensor elements are found to closely follow their
average behaviours, equations (\ref{diff:tensorD11}), (\ref{diff:tensorD22})
and (\ref{diff:tensorD33}).

In conclusion of this section, we have shown that
the classical dynamics of the model Hamiltonian (\ref{eq:KR3Dquasiper})
truly corresponds to a 3D chaotic diffusion, anisotropic along the true spatial direction {}``1''
and the {}``virtual'' directions {}``2'' and {}``3''. The comparison
of numerical simulations with the calculations of sect.~\ref{sec:AnalysisDynamics}
shows a very good agreement.

\section{Conclusion}

In this paper, we have reported some important statistical properties
of the classical transport in the 3D system corresponding to the
quasiperiodic kicked rotor we have recently
used to study the Anderson metal-insulator transition \cite{AP:Anderson:PRL08,AP:AndersonLong:PRA09}. The classical dynamics has been shown
to be fully diffusive in all directions, a prerequisite of the possibility of 3D Anderson localisation in the quantum regime.
We have given a simple analysis of the anisotropic character of the
transport, which is found
to be in good agreement with numerical
simulations. This should enable us to gain more insight into the metal-insulator phase diagram of the quasiperiodic kicked rotor \cite{Lemarie:critvsanisotropy}. 

\bibliographystyle{tMOP}

\begin{thebibliography}{33}
\providecommand{\natexlab}[1]{#1}

\bibitem[1]{Anderson:LocAnderson:PR58}
Anderson, P.W. {Absence of Diffusion in Certain Random Lattices}.  {\em Phys.
  Rev.}  {\bf 1958}, {\em 109} (5), 1492--1505.

\bibitem[2]{Chirikov:ChaosClassKR:PhysRep79}
Chirikov, B.V. {A universal instability of many-dimensional oscillator
  systems}.  {\em Phys. Rep.}  {\bf 1979}, {\em 52} (5), 263--379.

\bibitem[3]{Raizen:QKRFirst:PRL95}
Moore, F.L.; Robinson, J.C.; Bharucha, C.F.; Sundaram, B.; et~al. {Atom optics
  realization of the quantum {$\delta$}-kicked rotator}.  {\em Phys. Rev.
  Lett.}  {\bf 1995}, {\em 75} (25), 4598--4601.

\bibitem[4]{Casati:LocDynFirst:LNP79}
Casati, G.; Chirikov, B.V.; Ford, J.; et~al. {\em {Stochastic behavior of a
  quantum pendulum under periodic perturbation} }; Vol. ~93, {Springer-Verlag},
  {Berlin, Germany}, 1979; pp 334--352.

\bibitem[5]{Fishman:LocDynAnderson:PRA84}
Grempel, D.R.; Prange, R.E.; Fishman, S. {Quantum dynamics of a nonintegrable
  system}.  {\em Phys. Rev. A}  {\bf 1984}, {\em 29}, 1639--1647.

\bibitem[6]{Atland:SuperSymKR:PRL96}
Altland, A.; Zirnbauer, M.R. Field Theory of the Quantum Kicked Rotor.  {\em
  Phys. Rev. Lett.}  {\bf 1996}, {\em 77} (22) (Nov), 4536--4539.

\bibitem[7]{Atland:DiagKR:PRL93}
Altland, A. Diagrammatic approach to Anderson localization in the quantum
  kicked rotator.  {\em Phys. Rev. Lett.}  {\bf 1993}, {\em 71} (1) (Jul),
  69--72.

\bibitem[8]{Raizen:LDynNoise:PRL98}
Klappauf, B.G.; Oskay, W.H.; Steck, D.A.; et~al. {Observation of noise and
  dissipation effects on dynamical localization}.  {\em Phys. Rev. Lett.}  {\bf
  1998}, {\em 81} (6), 1203--1206.

\bibitem[9]{DArcy:AccModes:PRL99}
Oberthaler, M.K.; Godun, R.M.; d'Arcy, M.B.; Summy, G.S.; et~al. {Observation
  of quantum accelerated modes}.  {\em Phys. Rev. Lett.}  {\bf 1999}, {\em 83}
  (22), 4447--4451.

\bibitem[10]{Philips:QuantumResTalbot:PRL99}
Deng, L.; Hagley, E.W.; Denschlag, J.; Simsarian, J.E.; Edwards, M.; Clark,
  C.W.; Helmerson, K.; Rolston, S.L.; et~al. {Temporal, Matter-Wave-Dispersion
  Talbot Effect}.  {\em Phys. Rev. Lett.}  {\bf 1999}, {\em 83} (26),
  5407--5411.

\bibitem[11]{Monteiro:DoubleKick:PRL04}
Jones, P.H.; Stocklin, M.M.; Hur, G.; et~al. {Atoms in Double-{$\delta$}-Kicked
  Periodic Potentials: Chaos with Long-Range Correlations}.  {\em Phys. Rev.
  Lett.}  {\bf 2004}, {\em 93} (22), 223002.

\bibitem[12]{Renzoni:RatchetOptLat:PRL04}
Jones, P.H.; Goonasekera, M.; Renzoni, F. {Rectifying Fluctuations in an
  Optical Lattice}.  {\em Phys. Rev. Lett.}  {\bf 2004}, {\em 93} (7), 073904.

\bibitem[13]{Summy:QKAPhaseSpace:PRL06}
Behinaein, G.; Ramareddy, V.; Ahmadi, P.; et~al. {Exploring the Phase Space of
  the Quantum {delta}-Kicked Accelerator}.  {\em Phys. Rev. Lett.}  {\bf 2006},
  {\em 97} (24), 244101.

\bibitem[14]{Gong:HarperKR:PRA08}
Wang, J.; Gong, J. {Proposal of a cold-atom realization of quantum maps with
  Hofstadter's butterfly spectrum}.  {\em Phys. Rev. A}  {\bf 2008}, {\em 77}
  (3), 031405(R)--31408.

\bibitem[15]{GarciaGarcia:AndersonKR3D:PRE09}
Wang, J.; García-García, A.M. {Anderson transition in a three-dimensional
  kicked rotor}.  {\em Phys. Rev. E}  {\bf 2009}, {\em 79} (3), 036206.

\bibitem[16]{AP:Bicolor:PRL00}
Ringot, J.; Szriftgiser, P.; Garreau, J.C.; et~al. {Experimental Evidence of
  Dynamical Localization and Delocalization in a Quasiperiodic Driven System}.
  {\em Phys. Rev. Lett.}  {\bf 2000}, {\em 85} (13), 2741--2744.

\bibitem[17]{AP:SubFourier:PRL02}
Szriftgiser, P.; Ringot, J.; Delande, D.; et~al. {Observation of Sub-Fourier
  Resonances in a Quantum-Chaotic System}.  {\em Phys. Rev. Lett.}  {\bf 2002},
  {\em 89} (22), 224101.

\bibitem[18]{AP:Reversibility:PRL05}
Lignier, H.; Chab{\'e}, J.; Delande, D.; Garreau, J.C.; et~al. {Reversible
  Destruction of Dynamical Localization}.  {\em Phys. Rev. Lett.}  {\bf 2005},
  {\em 95} (23), 234101.

\bibitem[19]{AP:SubFMecs:EPL05}
Lignier, H.; Garreau, J.C.; Szriftgiser, P.; et~al. {Quantum diffusion in the
  quasiperiodic kicked rotor}.  {\em Europhys. Lett.}  {\bf 2005}, {\em 69}
  (3), 327--333.

\bibitem[20]{AP:PetitPic:PRL06}
Chab{\'e}, J.; Lignier, H.; Cavalcante, H.; Delande, D.; Szriftgiser, P.;
  et~al. {Quantum Scaling Laws in the Onset of Dynamical Delocalization}.  {\em
  Phys. Rev. Lett.}  {\bf 2006}, {\em 97} (26), 264101.

\bibitem[21]{Casati:IncommFreqsQKR:PRL89}
Casati, G.; Guarneri, I.; Shepelyansky, D.L. {Anderson transition in a
  one-dimensional system with three incommensurable frequencies}.  {\em Phys.
  Rev. Lett.}  {\bf 1989}, {\em 62} (4), 345--348.

\bibitem[22]{Basko:WeakDynLocQP:PRL03}
Basko, D.M.; Skvortsov, M.A.; Kravtsov, V.E. Dynamic Localization in Quantum
  Dots: Analytical Theory.  {\em Phys. Rev. Lett.}  {\bf 2003}, {\em 90} (9)
  (Mar), 096801.

\bibitem[23]{Lemarie:UnivAnderson:EPL09}
Lemari{\'e}, G.; Gr{\'e}maud, B.; Delande, D. {Universality of the Anderson
  transition with the quasiperiodic kicked rotor}.  {\em Europhys. Lett.}  {\bf
  2009}, {\em 87}, 37007.

\bibitem[24]{AP:Anderson:PRL08}
Chab{\'e}, J.; Lemari{\'e}, G.; Gr{\'e}maud, B.; Delande, D.; Szriftgiser, P.;
  et~al. {Experimental Observation of the Anderson Metal-Insulator Transition
  with Atomic Matter Waves}.  {\em Phys. Rev. Lett.}  {\bf 2008}, {\em 101}
  (25), 255702.

\bibitem[25]{AP:AndersonLong:PRA09}
Lemari{\'e}, G.; Chab{\'e}, J.; Szriftgiser, P.; Garreau, J.C.; Gr{\'e}maud,
  B.; et~al. {Observation of the Anderson metal-insulator transition with
  atomic matter waves: Theory and experiment}.  {\em Phys. Rev. A}  {\bf 2009},
  {\em 80} (4), 043626.

\bibitem[26]{Shepelyanky:Kq:PD87}
Shepelyansky, D.L. {Localization of diffusive excitation in multi-level
  systems}.  {\em Physica D}  {\bf 1987}, {\em 28}, 103--114.

\bibitem[27]{Maret:AndersonTransLight:PRL06}
St{\"o}rzer, M.; Gross, P.; Aegerter, C.M.; et~al. {Observation of the Critical
  Regime Near Anderson Localization of Light}.  {\em Phys. Rev. Lett.}  {\bf
  2006}, {\em 96} (6), 063904.

\bibitem[28]{Fishman:LocDynAnders:PRL82}
Fishman, S.; Grempel, D.R.; Prange, R.E. {Chaos, quantum recurrences, and
  Anderson localization}.  {\em Phys. Rev. Lett.}  {\bf 1982}, {\em 49},
  509--512.

\bibitem[29]{Leboeuf:LocBicPot:PRA10}
Albert, M.; Leboeuf, P. {Localization by bichromatic potentials versus Anderson
  localization}.  {\em Phys. Rev. A}  {\bf 2010}, {\em 81} (1),
  013614--Phys.Rev.A81,013614(2010)[8pages].

\bibitem[30]{Evangelou:LocDelocSolids:PRB94}
Panagiotides, N.A.; Evangelou, S.N.; Theodorou, G. {Localization-delocalization
  transition in anisotropic solids}.  {\em Phys. Rev. B}  {\bf 1994}, {\em 49}
  (20), 14122--14127.

\bibitem[31]{Shepelyanski:Bicolor:PD83}
Shepelyansky, D.L. {Some statistical properties of simple classically
  stochastic quantum systems}.  {\em Physica D}  {\bf 1983}, {\em 8}, 208--222.

\bibitem[32]{Rechester:KRDiffCoeff:PRA81}
Rechester, A.B.; Rosenbluth, M.N.; White, R.B. {Fourier-space paths applied to
  the calculation of diffusion for the Chirikov-Taylor model}.  {\em Phys. Rev.
  A}  {\bf 1981}, {\em 23} (5), 2664--2672.

\bibitem[33]{Lemarie:critvsanisotropy}
Lemari\'e, G.; Delande, D. {\em in preparation}  {\bf 2010}.

\end{thebibliography}

\end{document}